\documentclass[superscriptaddress,aps,prm,twocolumn,longbibliography]{revtex4-1}
\usepackage{graphicx}
\usepackage{epstopdf}
\usepackage{dcolumn}
\usepackage{bm}
\usepackage{color}
\usepackage{amsmath}
\usepackage{ulem}
\usepackage{graphicx}
\usepackage[colorlinks,linkcolor=blue, urlcolor=blue, anchorcolor=blue, citecolor=blue]{hyperref}

\begin{document}
\title{Epitaxial Growth and Band Structure of Antiferromagnetic Mott Insulator CeOI}

\author{Xinqiang Cai}
\affiliation{State Key Laboratory of Low-Dimensional Quantum Physics, Department of Physics, Tsinghua University, Beijing 100084, China}

\author{Zhilin Xu}
\affiliation{State Key Laboratory of Low-Dimensional Quantum Physics, Department of Physics, Tsinghua University, Beijing 100084, China}

\author{Hui Zhou}
\affiliation{Institute of Physics, Chinese Academy of Sciences, Beijing 100190, China}

\author{Jun Ren}
\affiliation{State Key Laboratory of Low-Dimensional Quantum Physics, Department of Physics, Tsinghua University, Beijing 100084, China}

\author{Na Li}
\affiliation{Key Laboratory for the Physics and Chemistry of Nanodevices, Department of Electronics, Peking University, Beijing 100871, China}

\author{Sheng Meng}
\email{smeng@iphy.ac.cn}
\affiliation{Institute of Physics, Chinese Academy of Sciences, Beijing 100190, China}

\author{Shuai-Hua Ji}
\email{shji@mail.tsinghua.edu.cn}
\affiliation{State Key Laboratory of Low-Dimensional Quantum Physics, Department of Physics, Tsinghua University, Beijing 100084, China}

\author{Xi Chen}
\email{xc@mail.tsinghua.edu.cn}
\affiliation{State Key Laboratory of Low-Dimensional Quantum Physics, Department of Physics, Tsinghua University, Beijing 100084, China}

\begin{abstract}
The van der Waals material CeOI is predicted to be a layered antiferromagnetic Mott insulator by DFT+U calculation. We successfully grow the CeOI films down to monolayer  on graphene/6H-SiC(0001) substrate by using molecular beam epitaxy. Films are studied  by {\it in-situ} scanning tunneling microscopy and spectroscopy, which shows a band gap of 4.4 eV. A metallic phase with composition unidentified also exists. This rare earth oxyhalide adds a new member to the two-dimensional magnetic materials.
\end{abstract}

%
%
%
%
%
\maketitle

\section{Introduction}
Two-dimensional (2D) magnetic materials  have triggered tremendous interest since the successful exfoliation of ferromagnetic (FM) Cr$_2$Ge$_2$Te$_6$ down to bilayer \cite{Cr2Ge2Te6nat} and CrI$_3$ down to monolayer \cite{CrI3nat}. Up to now, the ultra-thin films of many magnetic van der Waals (vdW) materials, such as FePS$_3$ \cite{FePS3nanolett, FePS32DMater}, CrI$_3$ \cite{CrI3nat}, CrBr$_3$ \cite{CrBr3natelectron}, Cr$_2$Ge$_2$Te$_6$ \cite{Cr2Ge2Te6nat, Cr2Ge2Te62DMater}, Fe$_2$GeTe$_3$ \cite{Fe3GeTe2nat}, VSe$_2$ \cite{VSe2PNAS, VSe2NatNanotech}, MnSe$_2$ \cite{MnSe2nanolett}, RuCl$_3$ \cite{RuCl3nanolett, RuCl3JPCS} and GdTe$_3$ \cite{GdTe3SciAdv}, have been created  by using adhesive tape, chemical vapor deposition (CVD) and molecular beam epitaxy (MBE).   These progresses  open a new avenue to exploring novel applications in spintronic devices at nanoscale.

In search of new vdW magnetic materials, metal oxyhalides emerge as a prototype with promising properties for practical application, such as robustness against air contamination. Monolayer transition metal oxyhalides, such as CrOBr and CrOCl, are recently predicted to be ferromagnetic insulators with antiferromagnetic bulk counterparts \cite{CrOXJACS}. Nevertheless, the magnetic property of rare earth oxyhalides has been rarely studied. In fact, the investigation of magnetic rare earth vdW materials is still at the early stage.

Here we study CeOI, whose material data are not available in literature yet. We predict layered CeOI as an antiferromagnetic  Mott insulator by first-principles calculation. Successful growth of CeOI down to monolayer by MBE is confirmed by X-ray photoelectron spectroscopy (XPS) and scanning tunneling microscopy (STM). With the help of scanning tunneling spectroscopy (STS), we reveal that CeOI film has  a band gap of 4.4 eV.

\section{Calculation}

The first-principles calculations of both bulk and monolayer CeOI based on density functional theory (DFT) were performed with Vienna Ab-initio Simulation Package (VASP) \cite{PRB.47.558, PRB.54.11169}. In the calculations, we employed  projector-augmented wave pseudopotential \cite{PRB.50.17953}, Perdew-Burke-Ernzerhof exchange-correlation functional (GGA) \cite{PRL.78.1396} and local density approximation (LDA) \cite{PRB.45.13244}. The energy cutoff of plane-wave basis is set as 520 eV. For the CeOI monolayer, the vacuum space is  larger than 15 {\AA}. The first Brillouin zone (BZ) is sampled using a Monkhorst-Pack scheme. For the monolayer (bulk), we use a k-mesh of $12\times12\times1$ ($6\times6\times6$) for structural optimization and $24\times24\times1$ ($12\times12\times12$) for the self-consistent calculations. The positions of atoms are optimized until the change of the total free energy is less than 0.0001 eV.  The convergence condition of electronic self-consistent loop is $10^{-5}$ eV. The GGA+U and LDA+U formalism formulated by Dudarev et al \cite{PRB.57.1505} have been used, since the plain GGA and LDA often fail to describe systems with localized f-electrons. We choose the effective U value ($U_{eff}=U-J$) as 5.8 eV for GGA+U and 6.8 eV for LDA+U, which are comparable to the $U_{eff}$ used in the calculation of cerium oxides \cite{PRB.75.035109, JCP.127.244704, JPCC.123.5164}. To investigate the magnetic property, spin-polarized calculations with collinear configurations were carried out.
\begin{figure}[htb]
\centering
\includegraphics[width=7cm] {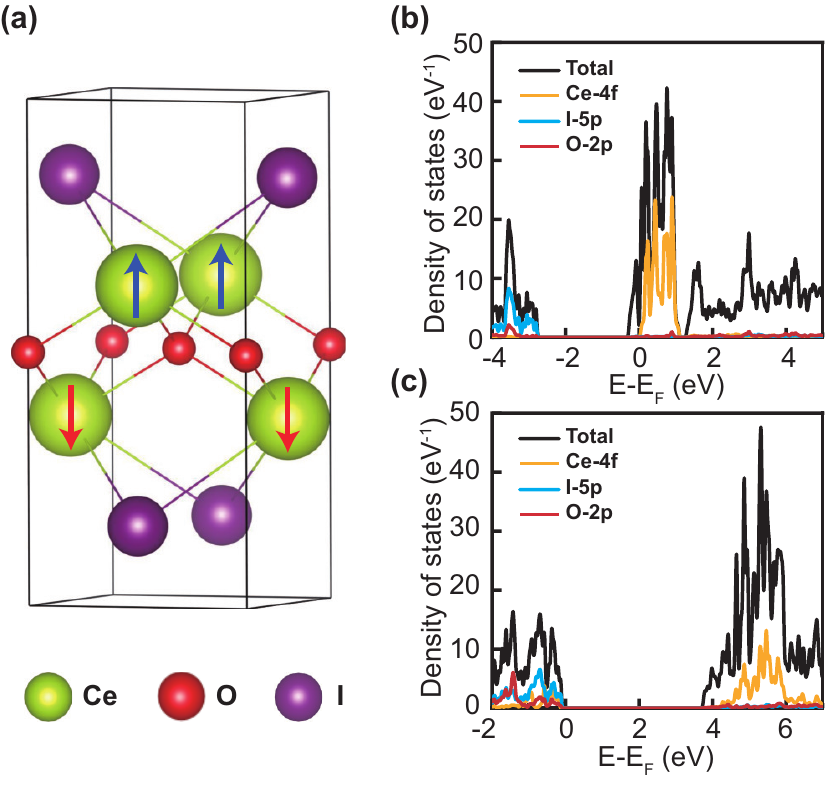}
\caption{(a) The crystal structure of CeOI. The directions of magnetic moments are illustrated by the blue and red arrows. (b) and (c) The calculated density of states of monolayer CeOI with $U_{eff}$ =0 eV and $U_{eff}$ =5.8 eV, respectively.
}
\label{}
\end{figure}

The bulk CeOI is a van der Waals bonded solid with tetragonal Bravais lattice and space group P4/nmm (No.129) (see Fig. 1(a)). The calculated crystal parameters are $a=b=4.156$ {\AA}, $c=9.656$ {\AA} for GGA+U and $a=b=4.070$ {\AA}. $c=8.872$ {\AA} for LDA+U, respectively.

For the CeOI monolayer, magnetism and band gap are estimated by  calculation using GGA+U.  The ferromagnetic  spin structure is 0.0172 eV/CeOI lower in energy for $U_{\rm{eff}}=0$ eV. But for larger $U_{\rm{eff}}$, the antiferromagnetic configuration is more stable, unveiling the antiferromagnetic ground state of CeOI. The calculated magnetic moment of CeOI per Ce atom is 0.990 $\mu_{B}$, indicating that there is one f-electron per Ce site in each unit cell which contains two Ce atoms. Moreover, the utilization of  $U$ introduces a metal (Fig. 1(b)) to insulator (Fig. 1(c)) transition in CeOI, showing that CeOI is a Mott insulator. The calculated band gap is 3.71 eV. As shown in Fig. 1(c), the lower Ce(4f) bands hybridize with O(2p) and I(5p) bands when $U_{\rm{eff}}=5.8$ eV, leading to possible underestimation of the band gap.

To verify the calculated properties, we performed effective $U$ test for monolayer CeOI.
The lattice parameter `$a$' as function of $U_{\rm{eff}}$ was calculated by different DFT approaches (see Fig. 2(a)). LDA slightly underestimates `$a$' while GGA always overestimates `$a$' by at least 0.01\AA. Between $U_{\rm{eff}}$=0 eV and $U_{\rm{eff}}$=1 eV, the curvature of `$a$' is more prominent than that for $U_{\rm{eff}}>$1 eV. The decrease in curvature corresponds to the separation of the occupied f-band from the unoccupied part. For $U_{\rm{eff}}>$1 eV, `$a$' assumes a constant approximately.

\begin{figure}[htb]
\centering
\includegraphics[width=5cm] {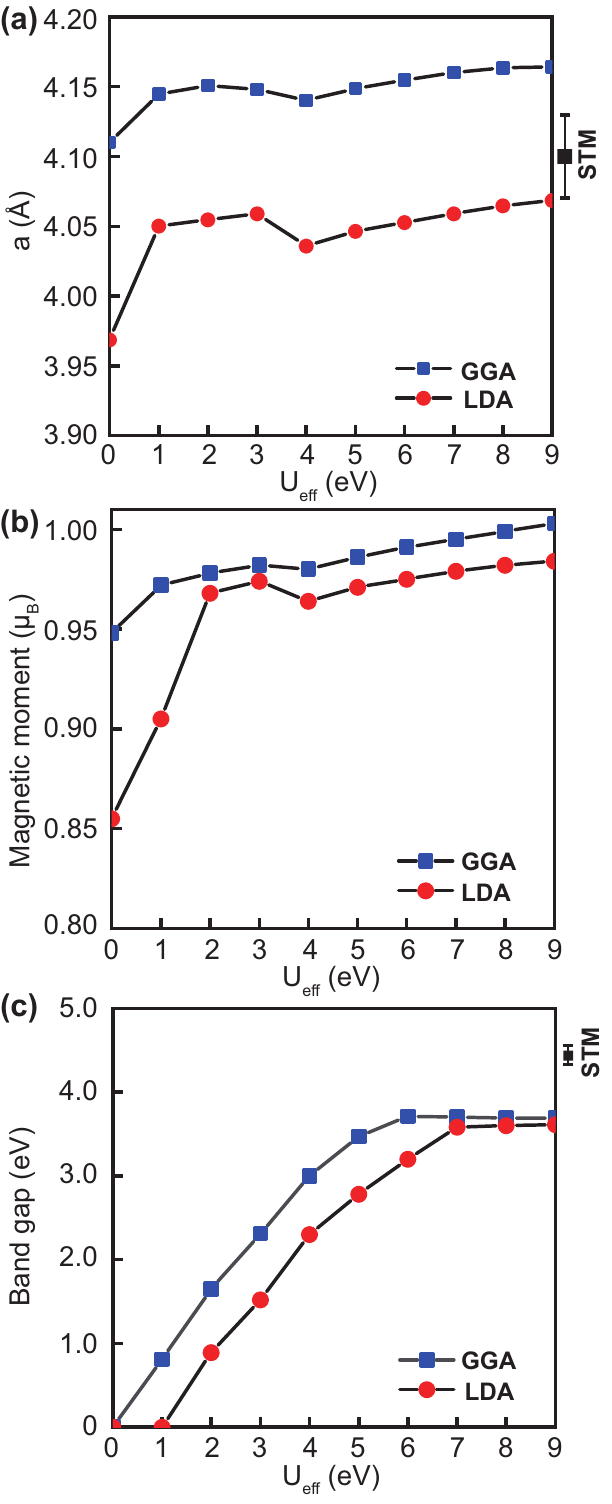}

\caption{(a), (b) and (c) show the lattice parameter `$a$', magnetic moment per Ce atom and band gap of CeOI monolayer as functions of $U_{\text{eff}}$, respectively. The black squares with error bars at the right side of (a) and (c) indicate the measured values by STM.
}
\label{}

\end{figure}

As shown in Fig. 2(b), the magnetic moment is fractional at low $U_{\rm{eff}}$ , indicating the itinerant character of the f-electron. But for $U_{\rm{eff}}>$2 eV, the f-electron is nearly completely localized in both GGA and LDA calculations and the magnetic moment approaches unity.

It should be noted that there is a dip in the vicinity of $U_{\rm{eff}}$=4 eV in both Fig. 2(a) and Fig. 2(b), probably originating from the onset of overlapping between the split lower Ce(4f) bands and the O(2p), I(5p) bands.

The band gap of CeOI monolayer increases linearly when $U_{\rm{eff}}$ increases from 0 eV to 4 eV in GGA and from 1 eV to 4 eV in LDA, while for $U_{\rm{eff}}>$ 4 eV the law of linear increment fails due to the interaction between the split lower Ce(4f) bands and the O(2p), I(5p) bands (Fig. 2(c)). It is assumed here that the distance between the conduction bands and the O(2p), I(5p) bands is big enough that no interactions exist.  The assumption is reasonable since DFT always underestimates the gap, i.e. the distance between the conduction bands and the O(2p), I(5p) bands.


\section{Experiments}

\subsection{Sample growth and characterization}

\begin{figure}[h]
\centering
\includegraphics[width=8cm] {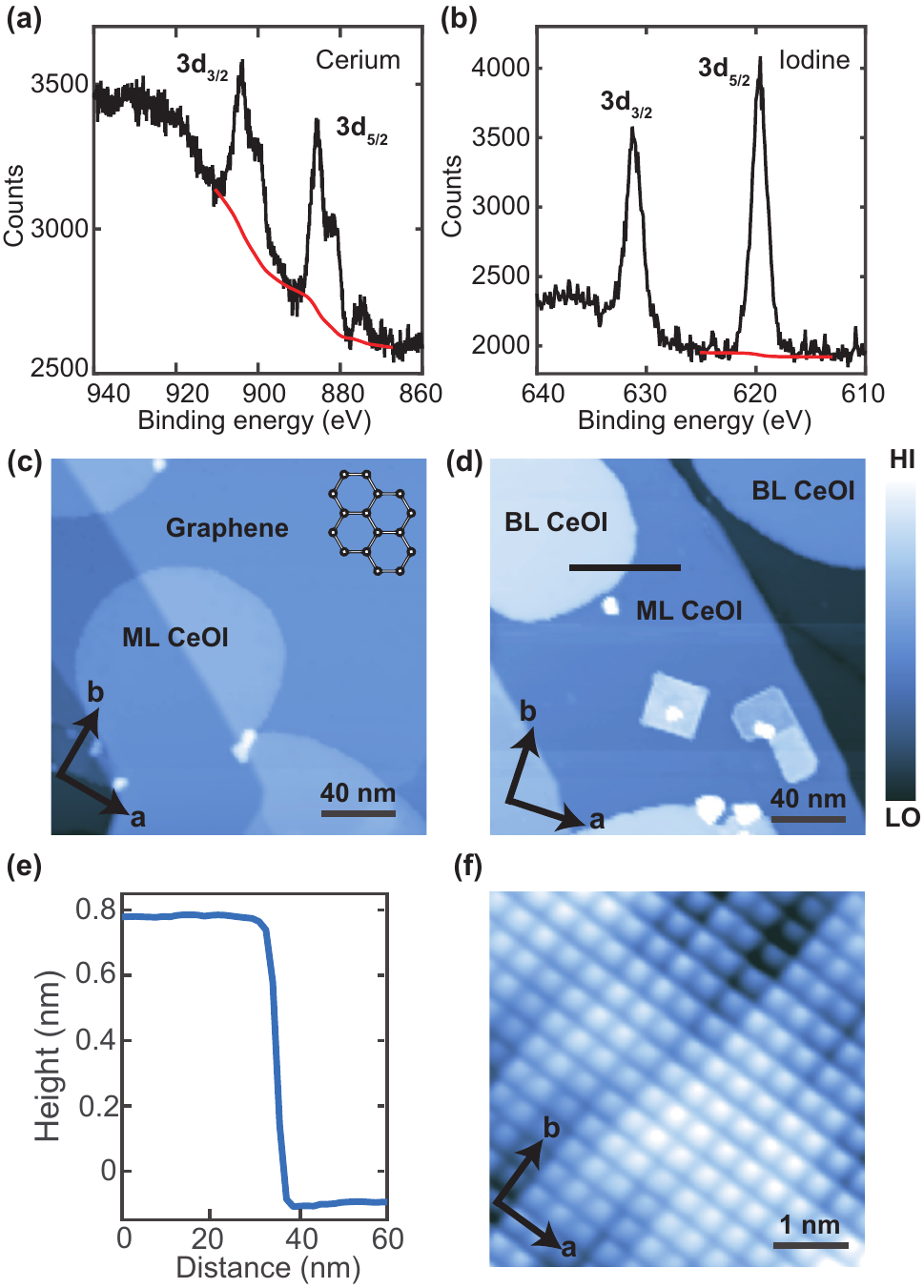}
\caption{(a) XPS of cerium (black curve) with background (red curve). (b) XPS of iodine (black curve) with background (red curve). (c) STM image of CeOI films on graphene with a coverage lower than one (bias: 4 V, current: 20 pA). The inset shows the lattice orientation of the graphene substrate. (d) STM image of CeOI films on graphene with a coverage between one and two (bias: 4 V, current: 20 pA). (e) Step height along the black line indicated in (d). (f) Atomic resolution of monolayer CeOI on graphene (bias: 0.3 V, current: 100 pA). Lattice constants are measured as $a=b=4.10\pm0.03 $ {\AA}.
}
\label{}
\end{figure}

We grew CeOI atomic layers via van der Waals epitaxy on single or bi-layer graphene prepared on a nitrogen-doped 6H-SiC(0001) substrate of resistivity below 0.1 $\Omega\cdot$cm by using the well-established recipe \cite{MBEGrapheneJPCM}. The base pressure of the MBE  system is about $1\times10^{-10}$ torr. The vacuum is better than $2\times10^{-9}$ torr during growth. Cerium flakes or foils (with purity of 99.5\%) were exposed to air and partially oxidized mainly into CeO$_2$ \cite{CeOxide} one day before introducing into the Knudsen cell on vacuum chamber. The cerium and oxygen atoms are therefore supplied by the partially oxidized cerium during thin film growth. The temperature of the cerium source is  from 1120$^\circ$C to 1180$^\circ$C. Under this temperature, CeO$_2$ is decomposed into Ce$_2$O$_3$ and O$_2$\cite{CeO2Decomposition}. The iodine is co-evaporated by  heating VI$_{3}$ or CrI$_3$ (over 240$^\circ$C for VI$_3$ and over 230$^\circ$C for CrI$_3$), which decomposes into solid VI$_2$ (or CrI$_2$) and gaseous I$_2$ at the elevated temperature\cite{VI3IC, CrI3JACS}. The decomposing rate of solid tri-iodide should be high enough so that the nominal flux of I$_2$ is much higher than those of Ce and O. To obtain high quality CeOI films, the growth rate needs to be very low, which is usually about 0.5 layer per hour.  The substrate temperature is set between 550$^\circ$C and 800$^\circ$C. The films exhibit two phases depending on the substrate temperature. The high  temperature ($\sim$800$^\circ$C) favors the formation of CeOI, while an unidentified metallic phase dominates  at lower temperature (550$^\circ$C to 600$^\circ$C).

The composition of the film was analyzed by X-ray photoelectron spectroscopy (XPS). To avoid possible reaction with O$_2$ and H$_2$O in air, the sample has been covered with layers of silicon before being transferred from the growth chamber to the XPS system. The XPS signals of cerium and iodine elements are showed in Figs. 3(a) and (b). Since the XPS is obtained {\it ex-situ}, the signal of oxygen is not relevant for making meaningful analysis. The shape and position of the XPS peaks for Ce element  are found to be the same as those of Ce$_2$O$_3$ \cite{Ce2O3JPCM}, confirming +3 valence for Ce. The valence of iodine atoms is found to be -1 by XPS \cite{CuIXPS}, where the main peak positions of $3d_{5/2}$ and $3d_{3/2}$ states are located at 619.6 eV and 631.3 eV, respectively. After eliminating the background by Shirley curve fitting (see red curves in Figs. 3(a) and (b)), cerium and iodine have an atomic concentration ratio of roughly 1:1. The XPS analysis agrees  with the chemical formula of CeOI.

The crystal and electronic structures of CeOI were analyzed by {\it in situ} STM. The sample was  transferred to the low-temperature STM (4.4 K) from the growth chamber without breaking the vacuum. Before imaging, the Pr-Ir alloy tips were checked and modified on the surface of Ag(111) islands. Figures 3(c) and (d) show the STM topography of typical samples with the coverage below 1 (Fig. 3(c)) and 1$\sim$2 (Fig. 3(d)), which demonstrates the layered structure and the layer-by-layer growth mode of CeOI. At initial stage of forming a new layer,  the adsorbed atoms usually crystalize into small rectangular grains. The grains further merge and form islands of various shapes. The step height of one van der Waals layer is measured to be 8.9 {\AA} (Fig. 3(e)), which is in agreement with calculation. The atomic resolution (Fig. 3(f)) can be achieved with a sample bias below 3 V. The a-axis of the lattice is preferentially aligned with one of the primitive lattice vectors of graphene. The in-plane square structure has a lattice constant of $a=b=4.10\pm0.03$ {\AA}, which also agrees well with calculation. It is interesting to point out that the lattice constant of CeOI is close to the bulk LaOI (4.14 {\AA}) and PrOI (4.09 {\AA}). The lattice constant data from STM combined with the valence analysis from XPS and first-principles calculations confirm the successful growth of CeOI film.

\subsection{STS spectra}

\begin{figure}[htb]
\centering
\includegraphics[width=8cm] {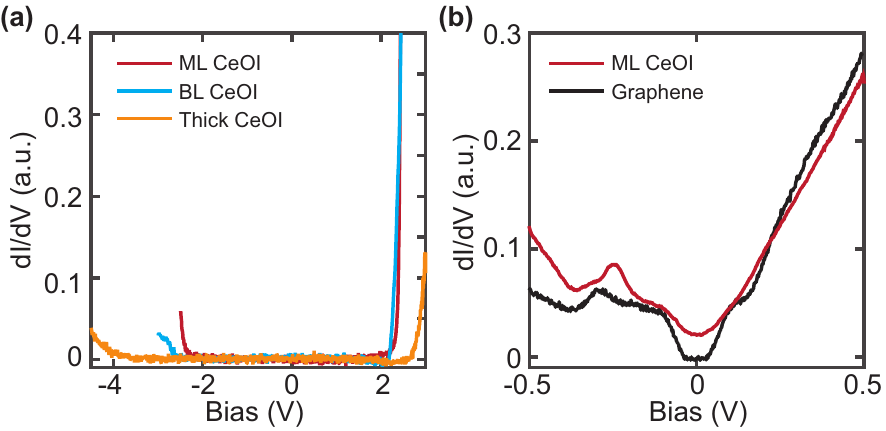}
\caption{(a) STS spectra of CeOI for monolayer and bilayer (set point for both: 2.5 V and 100 pA) together with thick films (set point: 3.2 V and 100 pA). (b)  STS spectra on monolayer CeOI  and  bilayer graphene (set point for both: 0.5 V and 100 pA).  }
\label{}
\end{figure}

The dI/dV spectra of STM measure the local density of states (LDOS) of a sample. In the experiment, the spectra were obtained through a lock-in amplifier with bias modulation of 5 mV at 913 Hz. In Fig. 4(a), the LDOS of single and double layers CeOI shows a band gap of $4.4\pm0.1$ eV. The deviation from  calculation  can be attributed to both the underestimate of band gap by DFT and the overestimate in STS measurement on insulating film. The latter effect, which is due to the partial voltage drop across the insulating layer, becomes more prominent for thicker films (over five layers CeOI).

The graphene substrate can also be detected through the CeOI monolayer ($\sim0.9$ \AA\ thick) if we choose a set point to bring the STM tip very close to the surface. The weak but finite  density of states (red curve in Fig. 4(b)) inside the band gap of CeOI is the remnant of the graphene LDOS (black curve in Fig. 4(b)). For bilayer CeOI film and beyond, the contribution of graphene to the LDOS at the STM tip apex is negligible and dI/dV is flat inside the band gap.

\begin{figure}[htb]
\centering
\includegraphics[width=6cm] {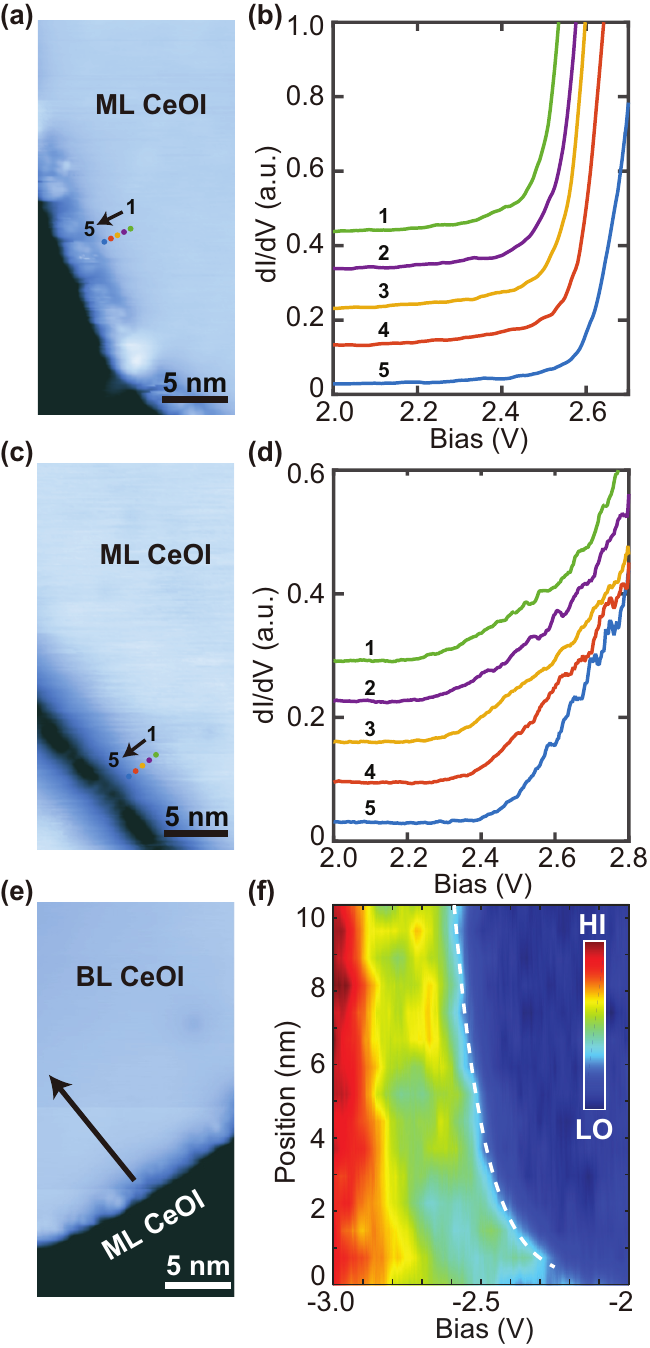}

\caption{(a) STM image (bias: 3 V, current: 20 pA) of monolayer CeOI near an step edge. (b) STS acquired on five points indicated in (a). Spectra are shifted vertically for clarity. (c) STM image (bias: 4 V, current: 20 pA) of monolayer CeOI near a grain boundary. (d) STS acquired on five points indicated in (c). Spectra are shifted vertically for clarity. (e) STM image (bias: 4 V, current: 20 pA) near a step between bilayer and monolayer CeOI. (f) Band bending near the step edge along the arrow in (e).
}
\label{}

\end{figure}

The charges trapped at defect sites make it possible to study the electrostatic interaction in this 2D material. Similar to another class of layered material transition metal dichalcogenides\cite{WSe22DMater}, band-bending is induced by the charged defects at the edge of CeOI films. The electrostatic potential generated by the charges locally shifts the band in energy. By comparing the onsets of conduction or valence band on the edge and in the interior of the film, the STS spectra (Fig. 5) acquired at the vicinity of edges provide a direct estimation of the shift. The band-bendings at the edge of monolayer (Figs. 5(a) and (b)), grain boundary of monolayer (Figs. 5(c) and (d)), and edge between bilayer  and monolayer CeOI (Figs. 5(e) and (f)) are found to be 0.18 eV, 0.15 eV, and 0.4 eV, respectively.

The upward shift towards the edge means the edge is negatively charged, which can be ascribed to the formation of dangling bonds around Ce vacancies on the edge. A very rough estimation of the linear charge density $\lambda$ is given by fitting the onset of valence band (dashed line in Fig. 5(f)) with the electrostatic potential as a function of the distance $r$ from the edge:
\begin{equation}
V(r)=\frac{\lambda}{2\pi\epsilon\epsilon_0}\ln\frac{r}{r_0}
\end{equation}
where the cutoff $r_0$ is set to one lattice constant. All the screening effects are taken account collectively by setting $\epsilon\sim10$. The charge density extracted from fitting is about 0.15 $e$ per unit cell, where $e=-1.6\times10^{-19} \text{ C}$.

\subsection{Coexisting metallic phase}

\begin{figure*}[htb]
\centering
\includegraphics[width=14cm] {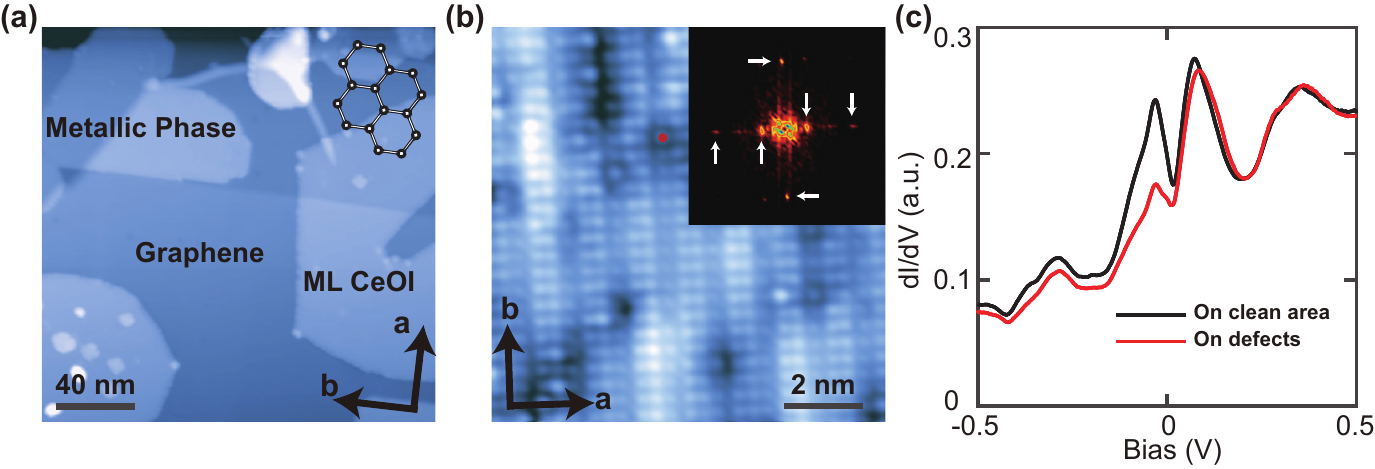}
\caption{(a) STM image of a film with both metallic phase and insulating CeOI (bias: 4 V, current: 20 pA). (b) Image of the metallic phase (bias: 0.5 V, current: 100 pA). Inset shows the FFT of  image with white arrows indicating the Bragg spots of the lattice and the ($3 {\times} 1$) superstructure.  (c) STS spectra on the metallic phase. }
\label{}
\end{figure*}

Besides the insulating CeOI, an orthogonal phase appears in the film as showed in Fig. 6(a) when the substrate temperature is below 600$^\circ$C. This phase is metallic and usually carries more defects than the insulating CeOI. The a-axis of the lattice is aligned with one of primitive lattice vectors of graphene. The lattice constants of this phase (see Fig. 6(b)) are $a=4.10\pm0.03$ {\AA} and $b=4.25\pm0.03$ {\AA}, respectively. A ($3 {\times} 1$) pattern can be seen in the atomically resolved image, which can be attributed to either the Moire pattern formed between the film and the graphene substrate or the intrinsic lattice superstructure. The STS (Fig. 6(c)) demonstrates the metallic behavior of such phase. The spectra (black curve of Fig. 6(c)) feature a dip in DOS at the Fermi level and two peaks at -30 meV and 70 meV, respectively.  The peak at -30 meV is weakened on a defect (red dot in Fig. 6(b)).  Except excluding  CeI$_2$ based on the lattice structure\cite{CeI2APA}, we are still not able to identify the composition of this metallic phase.

\section{Summary}

We calculated the band structure and magnetic configuration of the layered rare earth metal oxyhalide CeOI and the grew the film by MBE. It is found that CeOI is a Mott insulator with antiferromagnetic spin structure by calculation. The lattice parameters calculated by GGA+U and LDA+U are confirmed by STM measurement. The band gap of CeOI is found to be 4.4 eV by STS spectra and is a little larger than the calculated 3.71 eV. Charged defects exist on the film edge and induce band bending.  The magnetic properties need to be revealed by further experiments.

\section{Acknowledgement}
This work is supported by Science Challenge Project (No. TZ2016004), the National Natural Science Foundation of China (Nos. 11874233 and 91850120) and the Ministry of Science and Technology of China (No. 2018YFA0305603).

\bibliography{CeOI}

\end{document}